\documentclass[10pt, twocolumn]{IEEEtran}

\makeatletter
\def\ifundefined{\@ifundefined}
\makeatother

\usepackage{setspace}
\usepackage{srcltx}  % this is for only pc version to track the dvi and winedt editor
\usepackage{amsmath}
\usepackage{amssymb}
\usepackage{graphics}
\usepackage{cite}
\usepackage{psfrag}
\usepackage{epsfig}
\usepackage{subfigure}

\newtheorem{thm}{Theorem}
\newtheorem{cor}{Corollary}

\newtheorem{lem}{Lemma}

\begin{document}
\renewcommand{\textfraction}{0}

\title{A Practical Approach to Polar Codes}

%\vspace{-1in}

\author{A. Eslami and H. Pishro-Nik
\thanks{This work was supported by the National Science
Foundation under grants CCF-0830614 and ECCS-0636569.}
\thanks{The authors are with the
Electrical and Computer Engineering Department, University of Massachusetts, Amherst, MA, USA ~(email:eslami, pishro@ecs.umass.edu).}}

\maketitle
\thispagestyle{empty}
%\vspace{-.4 in}
\begin{abstract}
%\vspace{-.1 in}
In this paper, we study polar codes from a practical point of view. In particular, we study concatenated polar codes and rate-compatible polar codes. First, we propose a concatenation scheme including polar codes and Low-Density Parity-Check (LDPC) codes. We will show that our proposed scheme outperforms conventional concatenation schemes formed by LDPC and Reed-Solomon (RS) codes. We then study two rate-compatible coding schemes using polar codes. We will see that polar codes can be designed as universally capacity achieving rate-compatible codes over a set of physically degraded channels. We also study the effect of puncturing on polar codes to design rate-compatible codes.
\end{abstract}

\normalsize
%\vspace{-.1 in}

\begin{keywords}
%\vspace{-.1 in}
Polar Codes, Rate-Compatible Codes, Concatenated Codes.
\end{keywords}

%\vspace{-.1 in}

\section{Introduction}
Polar codes, recently proposed by Ar{\i}kan, are the first class of provably capacity achieving  codes for symmetric binary-input discrete memoryless channels (BDMC) with low encoding and decoding complexity \cite{arikan09}. Since Arikan's paper, there have been many papers studying the performance characteristics of polar codes (see for example \cite{arikan08, hussami09, korada09, koradait10,  mahdavifar10, bakshi10, Hof10}).
However, the research on polar codes, to a great extent, has been limited to the theoretical issues.
In this paper, we will look at two important practical problems that can be addressed using polar codes: concatenated codes, and rate-compatible codes. As it is shown, polar codes prove to be very powerful in both applications.
In addition to the capacity-achieving capability, polar codes have several other interesting properties from the practical point of view (such as having good error floor performance \cite{eslamiallerton10}). This suggests that a combination of polar coding with another coding scheme could eliminate shortcomings of both, and provide a powerful coding paradigm.
In this paper, we consider the design of polar code-based concatenated coding schemes that can contribute to closing the gap to the capacity. Since its introduction by Dave Forney in 1966, concatenated coding has been studied extensively for different combinations
of coding schemes and different methods of concatenations have been proposed. Furthermore, there
have been many applications, such as deep space communications, magnetic recording channels, and optical transport systems that use a concatenated coding scheme \cite{kurtas06, wu10, ningde08, Mizuochi09}.
A coding scheme employed in these applications needs to show strong error correction capability. In optical fiber communications, a minimum bit error rate (BER) of at least $10^{-12}$ is generally required \cite{ningde08,Mizuochi09}. Here, we investigate the potentials of using polar codes in a concatenated scheme to achieve such low error rates without error floor.
The problem of designing concatenated schemes using polar codes is yet to be
studied. The only previous work on this topic is \cite{bakshi10} where the authors study the classical idea of
code concatenation using short polar codes as inner codes and a high-rate Reed-Solomon (RS) code
as the outer code. It is shown that such a concatenation scheme with a careful choice of parameters
boosts the rate of decay of the probability of error to almost exponential in the block-length with
essentially no loss in computational complexity. While \cite{bakshi10} mainly considers the asymptotic case, we are interested in improving the performance in practical finite lengths.

In this paper, we study practical concatenation schemes that achieve a performance
close to capacity, while keeping a low complexity and avoiding error floor. We study the
combination of polar codes with LDPC codes, assuming polar codes as the outer code and LDPC
codes as the inner code. Comparing our proposed scheme against conventional concatenated schemes which employ a concatenation of RS, BCH, or LDPC codes we will show that polar-LDPC combination can actually outperform existing schemes. Our results suggest that polar codes have a great potential to be used in combination with other codes in real-world communication systems.

Rate-compatibility over time-varying channels is another important practical issue where error-correction codes are required to
be flexible with respect to their code rates depending on the current channel state.
In this paper, we study polar codes for rate-compatible applications.
We first see that polar codes are inherently well-suited for rate-compatible applications. We present a simple rate-compatible scheme that can universally achieve the channel capacity for a set of channels, using the same encoder and decoder. We will then study puncturing to design rate-compatible polar codes. Puncturing is widely used in the literature to generate rate-compatible codes \cite{ha04, ha06, hosseinitpuncture}. We will investigate the performance of random puncturing and stopping-tree puncturing (explained later) as used for polar codes and compare them to the universally capacity-achieving scheme. The rest of this paper is organized as follows. We first explain the notations and provide a short background on polar codes. Then, we study Concatenated polar codes  in Section \ref{sec:concat}. Rate-compatible polar codes are studied in Section \ref{sec:ratecomp}. Finally, Section \ref{sec:conclusion} concludes the paper.

\section{Preliminaries}\label{sec:pre}
In this section, we explain some notations and preliminary concepts we will be using in our analysis.
The construction of polar codes is based on the following
observation: Let $F = \left[ \begin{smallmatrix} 1&0\\ 1&1 \end{smallmatrix} \right]$.
%\begin{align}
%F= \[\left( \begin{array}{cc} 1 & 0 \\ 1 & 1 \end{array} \right)\].
%\end{align}
Apply the transform $F^{\otimes n}$ (where $\otimes n$ denotes the $n$th Kronecker power) to a block
of $N = 2^n$ bits and transmit the output through independent
copies of a symmetric BDMC, call it $W$. As $n$ grows large,
the channels seen by individual bits (suitably defined in \cite{arikan09})
start polarizing: they approach either a noiseless channel or a
pure-noise channel, where the fraction of channels becoming
noiseless is close to the capacity $I(W)$.
The channel polarization phenomenon suggests to use the
noiseless channels for transmitting information while fixing
the symbols transmitted through the noisy ones to a value
known both to the sender and receiver. Since the fraction of channels
becoming noiseless tends to $I(W)$, this scheme achieves the
capacity of the channel. In the following, let
$\bar{u}=(u_1, . . . , u_N)$, $\bar{x}=(x_1, . . . , x_N)$, and $\bar{y}=(y_1, . . . , y_N)$ denote, respectively, the vectors of input bits, code-bits, and channel output bits.
We denote the channels seen by input bits by $W^{(i)}$, $i=1, 2,..., N$ and call them bit-channels.

The generator matrix of polar codes, denoted by $G_N$, is a sub-matrix of the Kronecker product $F^{\otimes n}$ in which only a subset of rows of $F^{\otimes n}$ are present. An equivalent way of expressing this is to say that the codewords are of the form $\bar{x}=\bar{u}F^{\otimes n}$,
where the components $u_i$ of $\bar{u}$ corresponding to the ``frozen bits" are fixed to 0 and the
remaining components contain the ``information". Information bits, according to \cite{arikan09}, are chosen to be the ones corresponding to the best (least noisy) polarized channels.
We denote the set of information bits by $\mathcal{A}_N(W)$.

\begin{figure}[t]
\centering
{\includegraphics[width =3 in , height=2.2 in]{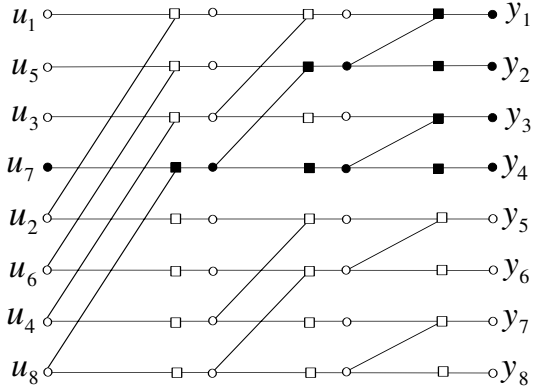}}
\caption{Normal realization of the encoding graph for $N=8$. An example of the stopping set is shown with black variable and check nodes.}
\label{fig:stopset1}
\end{figure}

A Successive Cancelation (SC) decoding scheme is employed in \cite{arikan09} to prove the capacity-achieving property of the code. However, \cite{arikan08} and \cite{hussami09} later employed belief propagation decoding to obtain better BER performance while keeping the decoding complexity at $O(N \log N)$.  Belief propagation can be run on the tanner graph of the code, easily obtained by adding check nodes to the encoding graph of polar codes; an example of such a graph is shown in Fig. \ref{fig:stopset1} for a code of length 8. We use a belief propagation decoder for the simulations throughout this paper.

In a rate-compatible coding scheme,
our goal is to provide reliable transmission over a set of channels with parameters $\theta^j , j = 1, ..., J$, using the same encoder and decoder.
Let $W_j$ and $c(\theta^j)$ denote a channel with parameter $\theta^j$ and its capacity, respectively. Assume that $\theta^i < \theta^j$ and
$c(\theta^i) > c(\theta^j)$ for $i > j$. We call a rate-compatible scheme
\emph{universally capacity achieving} (UCA), if the sequences of codes generated according to that scheme achieve the channel capacity $c(\theta^j)$ for
$j = 1, .., J$.
Assume that if $\theta^i < \theta^j$ then $W_j$ is a degraded version of $W_i$, which we denote by $W_j\preceq W_i$.
We may also assume that the channel state
information is available at the transmitter so that the appropriate code rate can be chosen for communication.

%The degree distribution for LDPC codes is represented by a polynomial pair $(\lambda, \rho)$ \cite{Richardsondes01} where $\lambda(x):=\sum_{i=2}^{d_v}\lambda_ix^{i-1}$  ($\rho(x):=\sum_{i=2}^{d_c}\rho_ix^{i-1}$) specifies the variable (check) node degree distribution. More precisely, $\lambda_i$ ($\rho_i$) represents the fraction of edges emanating from variable (check) nodes of degree $i$. The maximum variable degree and check degree is denoted by $d_v$ and $d_c$, respectively.

\section{Concatenated Polar Codes}\label{sec:concat}
\begin{figure}[t]
\centering
{\includegraphics[width=3.5 in , height=1.3 in]{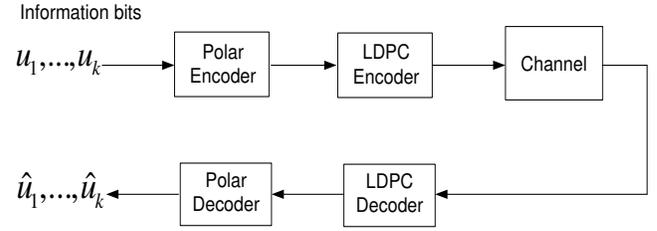}}
\caption{Block diagram of the proposed concatenated system of polar and LDPC codes.}
\label{fig:concat}
\end{figure}

In this section, we study concatenated polar codes. Our proposed scheme is formed of a Polar code as the outer code, and a LDPC code as the inner code. Fig. \ref{fig:concat} shows the block diagram of this scheme.
We consider long powerful LDPC codes as the inner code with rates close to the channel capacity.
LDPC codes can be decoded in linear time using BP, and at the same time can get very close to the capacity. However, LDPC codes with good waterfall characteristics are known to mostly suffer from the error floor problem.
Here, polar codes come to play their role.
In \cite{eslamiallerton10}, we studied the error floor performance of polar codes in finite length.
As it is shown in \cite{eslamiallerton10}, polar codes show very good error floor performance. Furthermore, the following theorem is proved in \cite{eslamiallerton10} about the girth of the tanner graph of polar codes.
\begin{thm}\label{th:girth}
Any polar code of length 8 or more has a girth of at least 12.
\end{thm}
Theorem \ref{th:girth} shows a natural advantage of polar codes over LDPC codes. In fact, a girth of 12 or more is considered so desirable for LDPC codes that many techniques have been proposed in the literature to guarantee such girths (see for example \cite{Esmaeili09} and references therein). Such a large girth also contributes to the good error floor performance.
Based on the facts mentioned above, the combination of polar and LDPC codes is expected to yield to a powerful concatenated code with BER performance close to the capacity for a broad range of the channel parameter.
Here we consider binary polar codes as the outer code concatenated with a binary LDPC code as the inner code. This is different from the traditional concatenated schemes \cite{lincostello83} in which a non-binary code is usually used as the outer code.

In order to investigate the performance of our proposed scheme, we consider a real-world application which uses a concatenated coding scheme as its essential part. An Optical Transport Network (OTN) is a set of optical network elements connected by optical fibre links, able to transport client signals at data rates as high as 100 Gbit/s and beyond. These networks are standardized under ITU-T Recommendation G.709 and are an important part of the high data-rate transmission systems such as Gigabit Ethernet and the intercontinental communication network. OTU4 is the standard designed to transport a 100 Gigabit Ethernet signal.
The FEC in the standard OTU4 employs a block interleaving of 16 words of the (255,239,17) Reed-Solomon codes, resulting in an overall overhead of 7\%. This scheme guarantees an error floor-free performance at least down to BERs of $10^{-15}$, and provides a coding gain of 5.8 dB at a BER of $10^{-13}$. Since the approval of this standard (February 2001), several concatenated coding schemes have been proposed in the literature and some as patents, targeting to improve the performance of this standard. In most cases, these schemes propose a concatenation of two of Reed-Solomon, LDPC, and BCH codes \cite{ningde08,wu10,Mizuochi09,Griesser}.
Here, for the first time, we consider polar-LDPC concatenation for the OTU4 setting.

\subsection{Encoder}
\begin{figure}[t]
\centering
{\includegraphics[width =3.5 in , height=2.5 in]{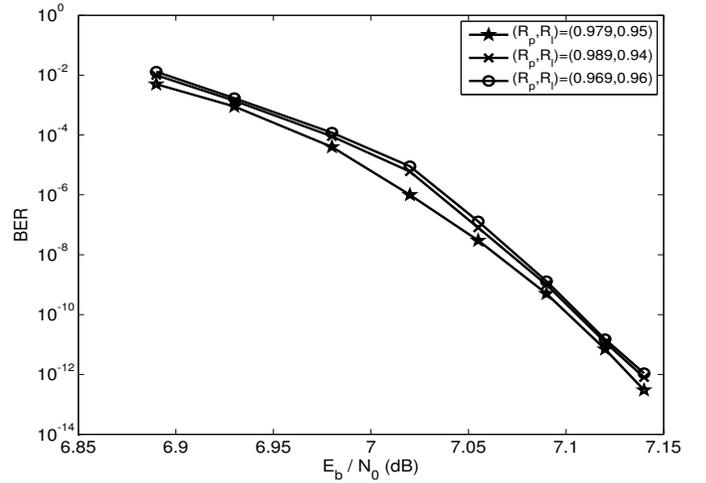}}
\caption{BER performance comparison for different rate combinations in a polar-LDPC concatenated scheme.}
\label{fig:ratecomb}
\end{figure}

In order to satisfy the overhead of 7\%, we adopt an effective code rate of 0.93. That is if we denote the code-rates for the polar and LDPC codes by $R_p$ and $R_l$ respectively, then $R_{eff}=R_p\times R_l$ needs to be 0.93.
The first problem is to find the optimal code-rate combination for the two codes to achieve the best BER performance. While this is an interesting analytical problem, it might be a difficult problem to solve. Therefore, we will find the best rate combination empirically. First, note that both $R_p$ and $R_l$ are greater than 0.93. We are also aware of the relatively poor error rate performance of finite-length polar codes compared to LDPC codes. Therefore, in order to minimize the rate loss, we choose $R_l$ close to the $R_{eff}$. As a result, $R_p$ would be close to 1. Now, we can use simulations to examine different code rates for $R_l$, each slightly above $R_{eff}$, to find the best choice. Fig. \ref{fig:ratecomb} shows the BER performance of three different rate couples, as a sample of all the rate couples we simulated. Code-length for the polar code is fixed to $2^{16}$ for all the rate couples. Showing a rate couple by $(R_p,R_l)$, these three rate couples are (0.989,0.94), (0.979,0.95), (0.969,0.96). We picked (0.979,0.95) for the rest of our simulations in this paper as it shows a better performance in the low-error-rate region and this is important for our application here. As it can be seen in the figure, choosing the rate couple can be a compromise for any specific application. Fixing the code-length $2^{16}$ for the polar code and fixing the rates to (0.979,0.95), the LDPC code-length would be 68985. We used the following optimal degree distribution pair which has a theoretical threshold of $0.435$:
\begin{align}
\nonumber \lambda(x)=0.502197 x+ 0.375436 x^2+ 0.061422 x^3+ 0.060945 x^9,
\end{align}
and
\begin{align}
\nonumber \rho(x)=0.076592 x^{34}+ 0.923408 x^{35}.
\end{align}

\subsection{Decoder}
At the decoder side, we perform belief propagation decoding with soft-decision for both the polar and LDPC codes. Upon finishing its decoding, the LDPC decoder will pass its output vector of LLRs to the polar decoder. Polar decoder then treats this vector as the input for its belief propagation process.

\subsection{Simulation Results}
\begin{figure}[t]
\centering
{\includegraphics[width =3.5 in , height=2.5 in]{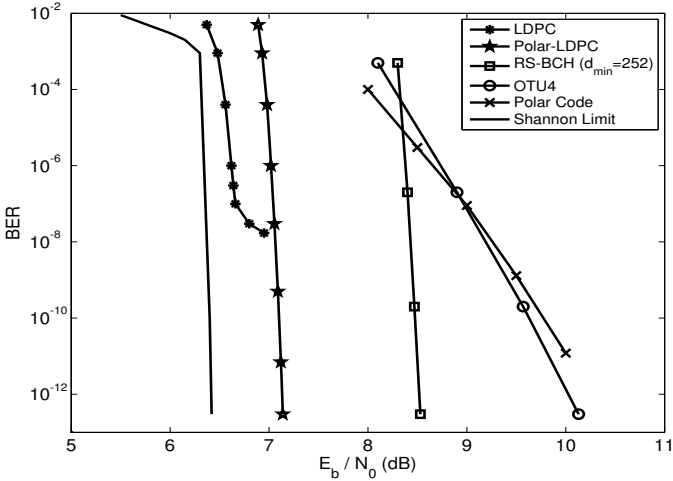}}
\caption{BER performance for different concatenated schemes.}
\label{fig:casc1}
\end{figure}

Fig. \ref{fig:casc1} depicts the BER performance for the concatenated scheme explained above, when using the LDPC code above. For the channel, we assumed a binary symmetric gaussian channel as it is used by \cite{ningde08,wu10,Mizuochi09,Griesser}.
Along with the concatenated scheme, we have shown the performance of the LDPC code and the polar code when used alone with an effective rate of 0.93, which is equal to the effective rate of the concatenated scheme.
As it can be seen, while the concatenated scheme shows a significant improvement over the polar code, it follows the performance of LDPC code in the waterfall region closely. Since both polar and LDPC codes here are capacity-achieving (or capacity-approaching), the concatenation does not suffer from rate loss in theory. Therefore, by increasing the code-length we expect that the curve for polar-LDPC scheme close the gap to the capacity. The curve also shows no sign of error floor down to BERs of $10^{-13}$, as opposed to the curve for the LDPC-only scheme.
In a polar-LDPC concatenation, the two codes are orchestrated to cover for each other's shortcomings as follows: LDPC plays the dominant role in its waterfall region, while polar code is dominant in the error floor region of the LDPC code.

To see the significant potential of polar codes for concatenated schemes, we also showed the BER performance corresponding to the G.709 standard explained earlier in the paper, as well as the scheme proposed in a recent patent \cite{Griesser}. Both systems use a code rate of 0.93. The scheme in \cite{Griesser} employs a RS-BCH concatenated code with an overall minimum distance of 252. The code-length for this scheme is 65280 which is very close to our proposed scheme. As it can be seen, at a BER of $10^{-12}$, polar-LDPC has an edge of 1.45 dB in SNR over the RS-BCH. An improvement of 2.9 dB over the G.709 standard can also be observed at this BER.

\section{Rate-Compatible Polar Codes}\label{sec:ratecomp}
In this section, we study two rate-compatible approaches to polar codes: universally capacity achieving rate-compatible polar codes, and punctured polar codes.
\subsection{Universally Capacity Achieving Rate-Compatible Polar Codes} \label{sec:uni rate}
Universally capacity achieving rate-compatible (UCARC) polar codes can achieve the channel capacity for a set of channels, using a low complexity encoder. UCARC polar codes can be explored using the following Lemma from \cite{satishthesis} (Lemma 4.7).

\begin{lem}\label{lem:satish}
Let $W$ and $\tilde{W}$ be two symmetric BDMCs such that $\tilde{W}$ is a degraded version of $W$. Then, $\tilde{W}^{(i)}$ is degraded with respect to $W^{(i)}$ for $i=1, ..., N$.
\end{lem}

This lemma implies that in a physically degraded setting, an order of polarization is maintained in the sense that ``good" bits for the degraded channel, must also be ``good" for the better channel. As a result, the set of information bits for the degraded channel is a subset of the set of information bits for the better channel, i.e. $\mathcal{A}_N(\tilde{W}) \subseteq \mathcal{A}_N(W)$.

\begin{cor}
Let $W_j,\ j=1,...,J$ be a set of symmetric BDMC channels such that $W_1\preceq W_2\preceq ...\preceq W_J$. Suppose that $\mathcal{A}_N(W_j)$ is known for $j=1,...,J$. Then, for any $i$ and $j$ such that $W_j\preceq W_i$, the capacity achieving polar code for $W_j$ can be obtained from the polar code designed for $W_i$, by setting the input bits in $\mathcal{A}_N(W_i) \setminus \mathcal{A}_N(W_j)$ to zero in the encoder.
\end{cor}

\begin{figure}[t]
\centering
{\includegraphics[width =2 in , height=4.5 in]{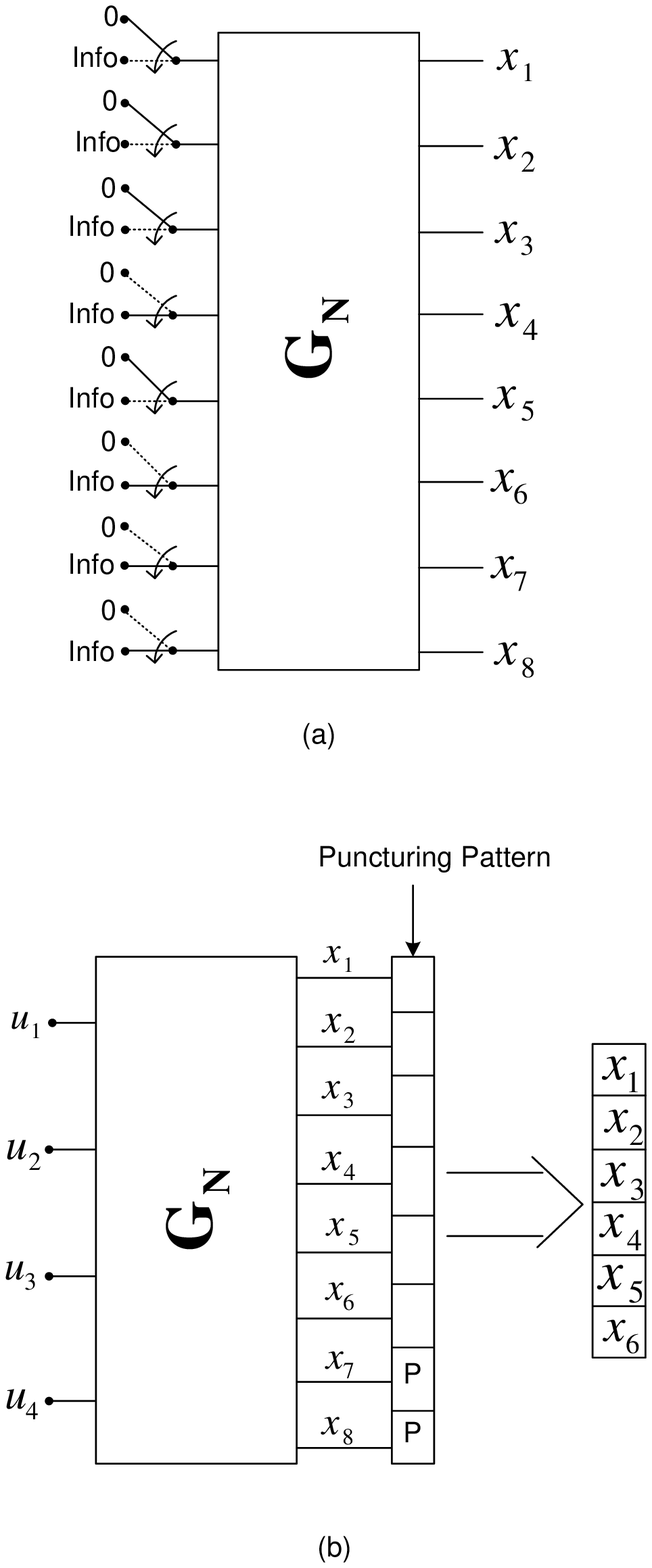}}
\vspace{-.1 in}
\caption{Different schemes for rate-compatible polar codes. (a) UCARC polar codes, (b) Punctured rate-compatible polar codes.}
\label{fig:schemes}
\vspace{-.1 in}
\end{figure}

This means that to implement different rates, the encoder only needs to shrink its set of information bits by switching a few of them to zero. This leads to a simple and practical structure for the UCARC polar codes.
This can be considered as an important advantage of polar codes over other coding schemes such as LDPC and turbo codes for which finding a UCARC scheme can be very complicated if even possible.
Fig. \ref{fig:schemes}(a) shows the structure of encoder for a UCARC polar code of length N=8. As it is shown, the input bits can be switched by the encoder to operate either as an information bit or a frozen bit.

\subsection{Puncturing for Rate-Compatible Polar Codes} \label{sec:punc}
In this section, we consider puncturing for rate-compatible polar codes. Figure \ref{fig:schemes}(b) shows the encoder structure for the punctured rate-compatible polar codes.
In this scheme, a parent code is designed for the worst channel (with largest channel parameter). In order to generate codes with higher rates for better channels, the encoder punctures some of the output bits. For every channel parameter $\theta^j, \ j=1,...,J$, a puncturing pattern is determined off-line and loaded into the encoder.
The punctured bits will not be sent over the channel. In the decoder side, the log-likelihood ratios for these bits will be set to zero before running belief propagation. The optimal puncturing pattern for each rate and a specific tanner graph can be found using optimization techniques \cite{ha04, ha06}; however, it turns out it is difficult to use such techniques for polar codes.

\subsubsection{Random Puncturing}\label{sec:randompunc}
\begin{figure}[t]
\centering
{\includegraphics[width =3.3 in , height=2.3 in]{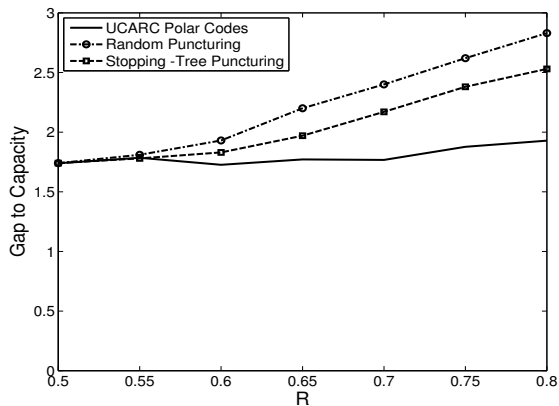}}
\vspace{-.1 in}
\caption{Gap-to-capacity for different rate-compatible schemes over the Gaussian channel. Parent-code rate for punctured codes is 1/2, parent code-length is $2^{13}$, and BER is fixed to $10^{-4}$.}
\label{fig:gap2cap}
\vspace{-.2 in}
\end{figure}

A simple way of puncturing, which is studied in some papers, is to have the encoder choose the punctured bits for each rate randomly.
Random puncturing is actually proved to be a UCARC scheme for LDPC codes over the BEC \cite{hosseinitpuncture}.
Fig. \ref{fig:gap2cap} shows the gap-to-capacity (for the gaussian channel) for randomly punctured polar codes compared against the UCARC polar codes described in section \ref{sec:uni rate}.
As it can be seen in the figure, there is a substantial distance between the two curves.

\subsubsection{Stopping-Tree Puncturing for Polar Codes}\label{sec:optimalpunc}
Here, we propose an algorithm involving the stopping sets in the tanner graph to improve the performance of puncturing.
A stopping set in the tanner graph is defined as a set of variable nodes such that every neighboring check node of the set is connected to at least two variable nodes in the set. Fig. \ref{fig:stopset1} shows an example of the stopping set in the polar codes' graph.
Stopping sets play an important role in the bit error rate and error floor performance of the code \cite{Urbankefinite02}.
A \emph{stopping tree} in the polar codes' tanner graph is a stopping set shaped as a tree rooted at an information bit (on the left side of graph) and with leaves at the code-bits (on the right side of graph), such as the one shown in Fig. \ref{fig:stopset1} with black variable and check nodes. We refer to such a tree as the stopping tree of an information bit. Stopping trees in fact form an important group of the stopping sets. We have shown in \cite{eslamiallerton10} that a minimum stopping tree is a minimum stopping set as well. For more on the structure of stopping sets and the importance of the stopping trees in belief propagation decoding of polar codes we refer the reader to \cite{eslamiallerton10}.

For any code-bit in the tanner graph, we can find the number of stopping trees that have that specific code-bit as a leaf node.
Then, we pick the punctured code-bits from the ones which are present in the fewest number of stopping trees.
This algorithm is based on the empirical results which show that the chance of recovery for these code-bits is higher than others in the case erasure. In other words, these code-bits are better protected than others in the tanner graph.
Since the information bits are known and the graph has a simple structure, we can easily find these bits. We call this algorithm \emph{Stopping-Tree Puncturing}.
As an example of this algorithm, suppose that we want to puncture the parent code of rate 1/2 in Fig. \ref{fig:schemes}(b) to a code of rate 3/4. Then we need to pick 2 code-bits to puncture. If we look at Fig. \ref{fig:stopset1}, we find out that $x_8$ is the only bit that is present in only one stopping tree. Among the code-bits which are present in two stopping trees we can choose $x_7$.

Fig. \ref{fig:gap2cap} shows the simulation results for stopping-tree puncturing compared to other techniques. As it can be seen, the gap-to-capacity has improved over the random puncturing though the distance to the UCARC scheme is still noticeable.

\section{Conclusion}\label{sec:conclusion}
We considered concatenated coding using polar codes. We proposed a polar-LDPC scheme and showed, via simulations, that it can result in considerable improvement over the existing concatenated schemes.
We also studied different approaches to rate-compatible polar codes over a set of physically degraded channels. We showed that UCARC polar codes can be designed with low complexity using the inherent characteristics of polar codes. We also studied the use of puncturing to generate rate-compatible polar codes.

{\footnotesize
\bibliographystyle{ieeetr}
%\linespread{1.6}
\bibliography{hldpcr,hldpcr1}

\begin{thebibliography}{10}

\bibitem{arikan09}
E.~Arikan, ``Channel polarization: A method for constructing capacity-achieving
  codes for symmetric binary-input memoryless channels,'' {\em IEEE
  Transactions o Information Theory}, vol.~55, pp.~3051--3073, July 2009.

\bibitem{arikan08}
E.~Arikan, ``A performance comparison of polar codes and reed-muller codes,''
  {\em IEEE Communications Letters}, vol.~12, no.~6, pp.~447 -- 449, 2008.

\bibitem{hussami09}
N.~Hussami, S.~Korada, and R.~Urbanke, ``Performance of polar codes for channel
  and source coding,'' in {\em IEEE International Sympousiom on Information
  Theory (ISIT)}, 2009.

\bibitem{korada09}
S.~Korada, E.~Sasoglu, and R.~Urbanke, ``Polar codes: Characterization of
  exponent, bounds, and constructions,'' in {\em IEEE International Symposium
  on Information Theory (ISIT)}, pp.~1483 -- 1487, 2009.

\bibitem{koradait10}
S.~Korada and R.~Urbanke, ``Polar codes are optimal for lossy source coding,''
  {\em IEEE Transactions on Information Theory}, vol.~56, no.~4, pp.~1751 --
  1768, 2010.

\bibitem{mahdavifar10}
H.~Mahdavifar and A.~Vardy, ``Achieving the secrecy capaity of wiretap channels
  using polar codes,'' in {\em IEEE International Symposium on Information
  Theory (ISIT)}, June 2010.

\bibitem{bakshi10}
M.~Bakshi, S.~Jaggi, and M.~Effros, ``Concatenated polar codes,'' in {\em IEEE
  International Symposium on Information Theory (ISIT)}, June 2010.

\bibitem{Hof10}
E.~Hof, I.~Sason, and S.~Shamai, ``Polar coding for reliable communications
  over parallel channels,'' in {\em IEEE Information Theory Workshop}, August
  2010.

\bibitem{eslamiallerton10}
A.~Eslami and H.~Pishro-Nik, ``On bit error rate performance of polar codes in
  finite regime,'' in {\em 48th Annual Allerton Conference on Communication,
  Control, and Computing}, August 2010.

\bibitem{kurtas06}
E.~M. Kurtas, A.~Kuznetsov, and I.~Djurdjevic, ``System perspectives for the
  application of structured {LDPC} codes to data storage devices,'' {\em IEEE
  Transactions on Magnetics}, vol.~42, no.~2, pp.~200 -- 207, 2006.

\bibitem{wu10}
C.~Wu and J.~Cruz, ``{RS} plus {LDPC} codes for perpendicular magnetic
  recording,'' {\em IEEE Transactions on Magnetics}, vol.~46, no.~16, pp.~1416
  -- 1419, 2010.

\bibitem{ningde08}
X.~Ningde, X.~Wei, Z.~Tong, E.~F. Haratsch, and M.~Jaekyun, ``Concatenated
  low-density parity-check and {BCH} coding system for magnetic recording read
  channel with 4 kb sector format,'' {\em IEEE Transactions on Magnetics},
  vol.~44, no.~12, pp.~4784 -- 4789, 2008.

\bibitem{Mizuochi09}
T.~Mizuochi, Y.~Konishi, Y.~Miyata, T.~Inoue, K.~Onohara, S.~Kametani,
  T.~Sugihara, K.~Kubo, H.~Yoshida, T.~Kobayashi, and T.~Ichikawa,
  ``Experimental demonstration of concatenated {LDPC} and {RS} codes by {FPGA}s
  emulation,'' {\em IEEE Photonics Technology Letters}, vol.~21, no.~18,
  pp.~1302 -- 1304, 2009.

\bibitem{ha04}
J.~Ha, J.~Kim, and S.~McLaughlin, ``Rate-compatible puncturing of low-density
  parity-check codes,'' {\em IEEE Transactions on Information Theory}, vol.~50,
  no.~11, pp.~2824--2836, 2004.

\bibitem{ha06}
J.~Ha, J.~Kim, and S.~McLaughlin, ``Rate-compatible punctured low-density
  parity-check codes with short block lengths,'' {\em IEEE Transactions on
  Information Theory}, vol.~52, no.~2, pp.~729--738, 2006.

\bibitem{hosseinitpuncture}
H.~Pishro-Nik and F.~Fekri, ``Results on punctured low-density parity-check
  codes and improved iterative decoding techniques,'' {\em IEEE Trans. on
  Inform. Theory}, vol.~53, pp.~599--614, February 2007.

\bibitem{Esmaeili09}
M.~Esmaeili and M.~Gholami, ``Geometrically-structured maximum-girth ldpc block
  and convolutional codes,'' {\em IEEE Journal on Selected Areas in
  Communications}, vol.~27, no.~6, pp.~831--845, 2009.

\bibitem{lincostello83}
S.~Lin and D.~J. Costello, {\em Error Control Coding: Fundamentals and
  Applications}.
\newblock Prentice-Hall, 1983.

\bibitem{Griesser}
H.~Griesser and J.~P. Elbers, ``Forward error correction coding.'' U.S. Patent,
  Jan 2009.
\newblock US 7,484,165 B2.

\bibitem{satishthesis}
S.~Korada, {\em Polar Codes for Channel and Source Coding}.
\newblock PhD thesis, Ecole Polytechnique Fédérale de Lausanne (EPFL), 2009.

\bibitem{Urbankefinite02}
C.~Di, D.~Proietti, I.~E. Telatar, T.~Richardson, and R.~Urbanke,
  ``Finite-length analysis of low-density parity-check codes on the binary
  erasure channel,'' {\em IEEE Trans. Inform. Theory}, vol.~48, pp.~1570
  --1579, 2002.

\end{thebibliography}

\end{document}